\newcommand{\beq}{\begin{equation}}
\newcommand{\eeq}{\end{equation}}
\newcommand{\bea}{\begin{eqnarray}}
\newcommand{\eea}{\end{eqnarray}}

\newcommand{\gsim}{\lower.7ex\hbox{$\;\stackrel{\textstyle>}{\sim}\;$}}
\newcommand{\lsim}{\lower.7ex\hbox{$\;\stackrel{\textstyle<}{\sim}\;$}}

\newcommand{\mrm}{\mathrm}

\documentclass[a4paper]{jpconf}
\usepackage{graphicx}
\usepackage{epstopdf}
\begin{document}
\title{Specific Dark Matter signatures from hidden U(1) }

\author{Yann Mambrini}

\address{ Laboratoire de Physique Th\'eorique,
Universit\'e Paris-Sud, F-91405 Orsay, France\\}

\ead{yann.mambrini@th.u-psud.fr}

\begin{abstract}
Several constructions motivate the existence of a dark
$U(1)_D$ gauge boson which interacts with the Standard Model only through its
kinetic mixing or loop induced processes. 
We describe two typical examples with specific signatures
 in particular we show that a region with relatively light WIMPS, 
$M_{Z_D} < 40$ GeV and a kinetic mixing $10^{-4} < \delta < 10^{-3} $
is not yet excluded by the last experimental data and seems to give 
promising signals in a near future. We also show that  conditions from anomaly cancelation 
 generate tri-vector couplings $Z_D Z \gamma$ leading to a specific gamma ray line observable
 by FERMI telescope.
\end{abstract}

\section{Introduction}
Neutral gauge sectors with an additional dark $U(1)_D$ symmetry in addition 
to the Standard Model (SM) hypercharge $U(1)_Y$ and an associated $Z_D$ 
are among the best motivated extensions of the SM, and give the possibility
that a dark matter candidate lies within this new gauge sector of the theory \cite{zprime}.
The new vector boson $Z_D$ can interact with the SM, even if no 
SM fermions are directly charged under the additional gauge symmetry.
This interaction can occurs via mixed kinetic terms between the SM's hypercharge
field strength and the new abelian field strength 
\cite{Feldman:2007wj,Mixing,Baumgart:2009tn,Chun:2010ve} 
or through couplings generated by counter-term to preserve the
anomaly cancelation condition \cite{Anastasopoulos:2006cz,Anomalies}. 
Whereas the former couplings can give significant signals in direct detection experiment
even fitting the last DAMA \cite{DAMA} or COGENT \cite{COGENT}   excesses
\cite{Andreas:2010dz,Mambrini:2010dq,Buckley:2010ve,Schwetz:2011xm} or INTEGRAL 511 keV line
\cite{Cline:2010kv}, the latter
can give rise to a gamma-ray line observable
 in satellite telescopes \cite{Dudas:2009uq,Mambrini:2009ad,Lines,Goodman:2010qn}.
 A summary of gamma-ray line constraints can be found in \cite{Gammasummary}, and within a
 supersymmetric \cite{Chen:2010kq} in
 extra-dimensional framework in \cite{Bertone:2010fn}. A light Z' can be justified in leptophobic constructions
 \cite{Gondolo:2011eq} and even be an interbretation of the Wjj anomaly observed by CDF \cite{Wjj}.
 Using LEP/Tevatron constraints, a higher dimensional approach can be found in \cite{Mambrini:2011pw}
 whereas a summary of all the constraints has been studied in \cite{Mambrini:2011dw}

\section{The dark kinetic mixing}

The matter content of any $dark$ $U(1)_D$ extension of the SM can be decomposed
into three families of particles:

\begin{itemize}
\item{The $Visible$ $sector$ is made of particles which are charged under the SM
gauge group $SU(3)\times SU(2)\times U(1)_Y$ but not charged under $U(1)_D$
(hence the $dark$ denomination for this gauge group)}
\item{the $Dark$ $sector$ is composed by the particles charged under
$U(1)_D$ but neutral with respect of the SM gauge symmetries. The dark matter
($\psi_0$) candidate is the lightest particle of the $dark$ $sector$}
\item{The $Hybrid$ $sector$ contains states with SM $and$ $U(1)_D$ quantum numbers. 
These states are fundamental because they act as a portal between
the two previous sector through the kinetic mixing they induce at loop
order.} 
\end{itemize}

\noindent
From these considerations, it is easy to build the effective lagrangian
generated at one loop :

\begin{eqnarray}
{\cal L}&=&{\cal L}_{\mrm{SM}}
-\frac{1}{4} \tilde B_{\mu \nu} \tilde B^{\mu \nu}
-\frac{1}{4} \tilde X_{\mu \nu} \tilde X^{\mu \nu}
-\frac{\delta}{2} \tilde B_{\mu \nu} \tilde X^{\mu \nu}
\nonumber
\\
&+&i\sum_i \psi_i \gamma^\mu D_\mu \psi_i
+i\sum_j \Psi_j \gamma^\mu D_\mu \Psi_j
\label{Kinetic}
\end{eqnarray}

\noindent
$B_{\mu}$ being the gauge field for the hypercharge, 
$X_{\mu}$ the gauge field of $U(1)_D$ and
$\psi_i$ the particles from the hidden sector, $\Psi_j$ the particles
 from the hybrid sector, 
$D_{\mu}  =\partial_\mu -i (q_Y \tilde g_Y \tilde B_{\mu} + q_D \tilde g_D
 \tilde X_{\mu} + g T^a W^a_{\mu})$, $T^a$ being the $SU(2)$ generators, and 

\beq
\delta= \frac{\tilde g_Y \tilde g_D}{16 \pi^2}\sum_j q_Y^j q_D^j 
\log \left( \frac{m_j^2}{M_j^2} \right)
\eeq

\noindent
with $m_j$ and $M_j$ being hybrid mass states \cite{Baumgart:2009tn}.
Notice that the sum is on all the hybrid states, as they are the only ones which can contribute to the $Y_{\mu}X_{\mu}$ propagator.
After diagonalization of the current eigenstates, one makes the gauge kinetic
terms of Eq.(\ref{Kinetic}) diagonal and canonical.

\begin{figure}
    \begin{center}
    \includegraphics[width=2.5in]{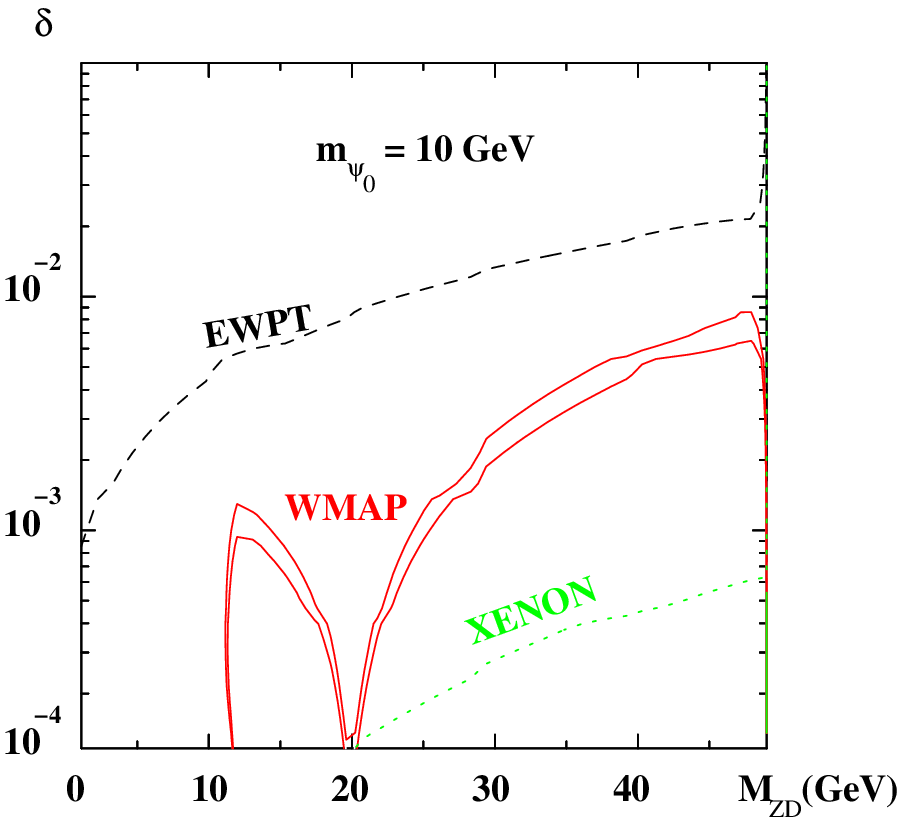}
    \hspace{0.3cm}
    \includegraphics[width=2.5in]{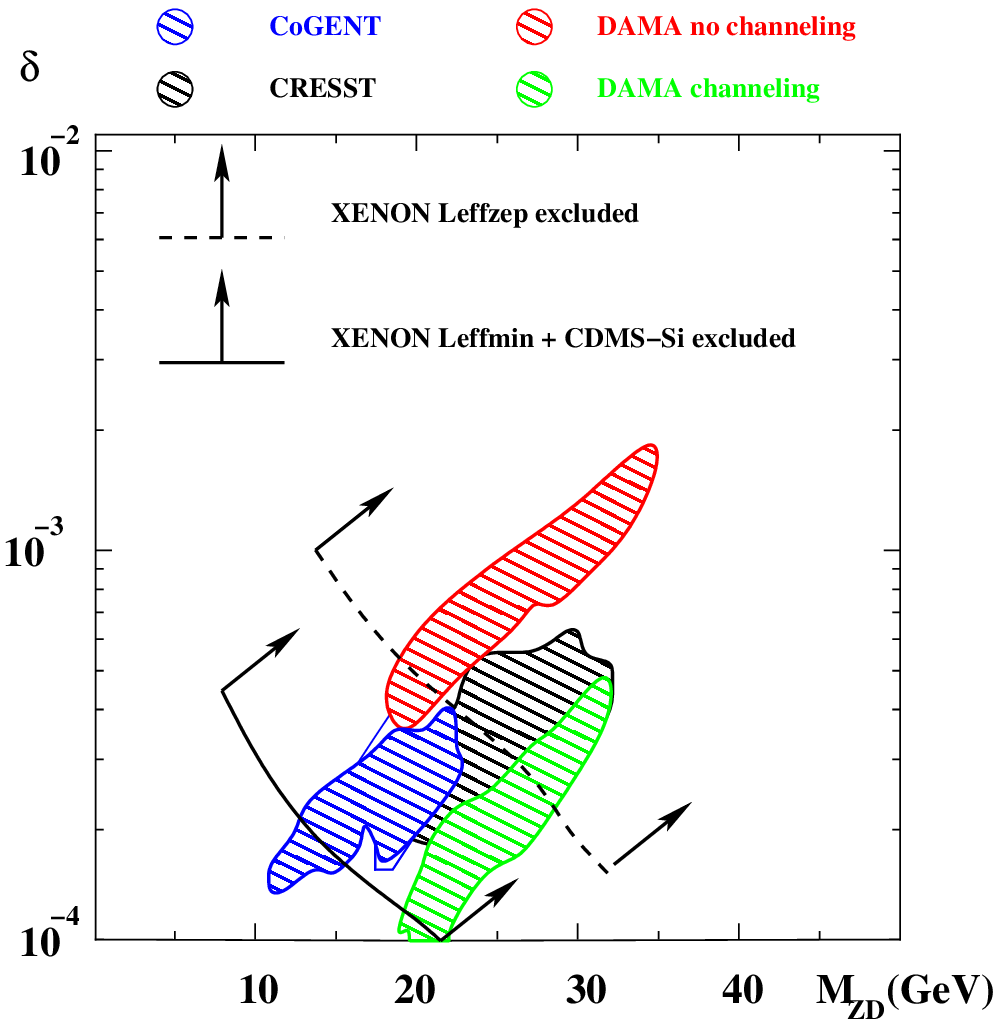}
          \caption{{\footnotesize
Left : example of allowed parameter space for $m_{\psi_0}=10$ GeV in the 
($M_{Z_D}$, $\delta$) plane (left). The points between the full-red region
respect the 5$\sigma$ WMAP constraint, the points below the dashed-black
line do not exceed accelerator data on precision tests, and the points
above the dotted-green line are excluded by XENON100 data. Right: 
parameter space allowed within 90 \% of  C.L. for the CoGeNT
signal (blue), DAMA without channeling (red), with channeling 
(green), CRESST (black), and the exclusion region depending on the hypothesis
concerning $L_{eff}$.
}}
\label{fig:Allowed}
\end{center}
\end{figure}

 We show in Fig.\ref{fig:Allowed} (left)
 the points that fulfill the WMAP $5\sigma$ bound \cite{WMAP}
 on $\mathrm{\Omega_{DM}}$ for $m_{\psi_0}= 10$ GeV
 in the ($M_{Z_D},\delta$) plane.
 One can clearly see the $Z_D-$pole region when 
 $M_{Z_D} \sim m_{\psi_0}$.
One important point is that for a given $M_{Z_D}$ and $m_{\psi_0}$, 
there exists a unique solution $\delta$ (up to the very small uncertainties at 5$\sigma$)
fulfilling WMAP constraints : from 3 parameters
($m_{\psi_0}, M_{Z_D}, \delta$), the WMAP constraints reduce it to two 
($M_{Z_D}, \delta$).

\noindent
 We show in Fig.\ref{fig:Allowed} the points respecting 
 WMAP, and the
 DAMA/LIBRA (with and without channeling) CoGeNT and 
 CRESST\footnote{For the CRESST estimation, we used an extrapolation 
 given in the talk of  T. Schwetz and the CRESST 
 collaboration \cite{CRESST}.} results at 90 \% of CL.
 All the constraints have been calculated for a standard Maxwellian velocity distribution
(with mean velocity 
$v_0=230$ km/s  and an escape velocity $v_{esc}=600$ km/s). 
 One can observe in Fig.\ref{fig:Allowed}  that for all experiments, the regions are quite surprisingly
 near and correspond to $10~\mrm{GeV} \lsim M_{Z_D} \lsim 30$ GeV
 and $10^{-4} \lsim \delta \lsim 10^{-3}$, which is in complete agreement
 with the measurement of electroweak precision tests. Moreover, such values
 of $\delta$ are typical of one loop-order corrections and can easily be
 generated by heavy-fermions loops in the $Z-Z_D$ propagator.
 
\section{Anomalies and gamma-ray line}

It is well known that any extension of the SM which introduces chiral fermions
with respect to gauge fields suffers from anomalies, a phenomenon of breaking of gauge
symmetries of the classical theory at one-loop level. Anomalies are responsible for instance for
a violation of unitarity and make a theory inconsistent \cite{ABJ,Coriano}.
For this reason if any construction introduces a new fermionic sector to address
the DM issue of the SM, it is vital to check the cancelation of anomalies
and its consequences on the Lagrangian and couplings.
The idea is to add to the Lagrangian local gauge 
non-invariant
terms in the effective action whose gauge variations cancel the anomalous triangle diagrams.
There exist two kinds of term which can cancel the mixed $U(1)_D\times G_A^{\mathrm{SM}}$
anomalies, with $G_A^{\mathrm{SM}}$ being one
of the SM gauge group $SU(3)\times SU(2)\times U_Y(1)$ :
the Chern Simons (CS) term which couples the $G_A^{\mathrm{SM}}$ to the 
$U(1)_D$  gauge boson,
and the Peccei-Quinn (PQ, or Wess-Zumino (WZ)) term which couples the 
$G_A^{\mathrm{SM}}$
gauge boson to an axion. In the effective action, these terms are sometimes called
Generalized Chern--Simons (GCS) terms \cite{Anastasopoulos:2006cz}:

\begin{eqnarray}
{\cal L}_{inv} &&= - \frac{1}{4 g'^2} F^{Y \mu \nu} F^Y_{\mu \nu}
- \frac{1}{4 g_X^2} F^{X \mu \nu} F^X_{\mu \nu}
-\frac{1}{2} (\partial_\mu a_X - M_X X_\mu)^2
-i \overline{\psi} \gamma^\mu D_\mu \psi
\nonumber
\\
{\cal L}_{var} &&=
\frac{C}{24 \pi^2} a_X \epsilon^{\mu \nu \rho \sigma} F^Y_{\mu \nu} F^Y_{\rho \sigma}
+\frac{E}{24 \pi^2} \epsilon^{\mu \nu \rho \sigma} X_{\mu} Y_{\nu} F^Y_{\rho \sigma}
\label{Lagrangian}.
\end{eqnarray}

\noindent
The Stueckelberg axion $a_X$ ensures the gauge invariance of the effective Lagrangian and
$g_X$ and $F^X_{\mu \nu}=\partial_{\mu} X_\nu - \partial_\nu X_\mu$
are the gauge coupling and field strength of $U(1)_D$. The axion
has a shift transformation under $U(1)_D$
\begin{equation}
\delta X_{\mu} \ = \ \partial_{\mu} \alpha \quad , \quad \delta a_X
\ = \ \alpha \  M_X .
\end{equation}

\noindent
The ${\cal L}_{var}$ will generate after the $SU(2)\times U(1)_Y$ tri-vectorial couplings
$Z_DZZ$ and $Z_D\gamma Z$ ($Z\gamma\gamma$ coupling being forbidden by spin-
momentum conservation). This generates new annihilation processes 
$\psi_0 \psi_0 \rightarrow Z_D \rightarrow ZZ/Z\gamma$ which can be observable through 
the $only$ monochromatic gamma--ray line with energy  
$E_\gamma= m_{\psi_0} \left( 1- \frac{M_Z^2}{4 m_{\psi_0}^2} \right)$
\cite{Dudas:2009uq,Mambrini:2009ad}. Other models predicts several lines
\cite{LineBuckley,Lines}, but none of them just one line.

\begin{figure}
    \begin{center}
    \hspace{-1cm}
    \includegraphics[width=2.5in]{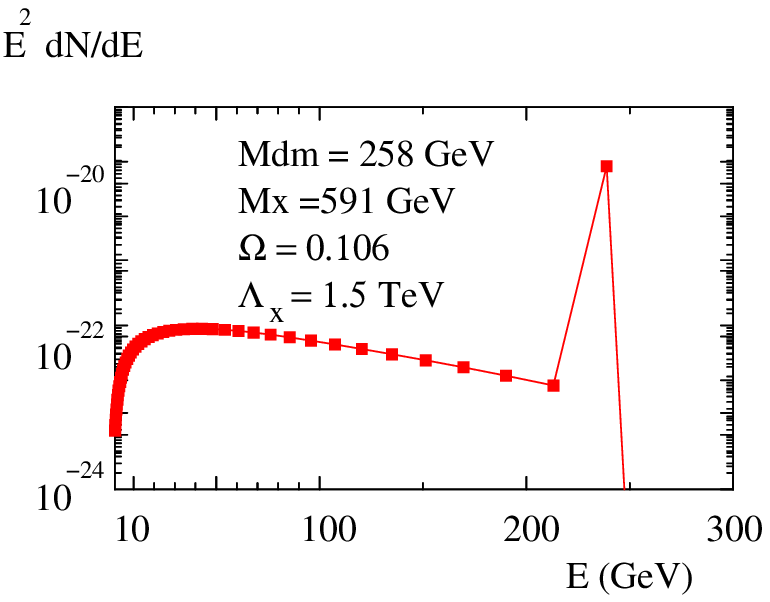}
    \hspace{1cm}
    \includegraphics[width=3.3in]{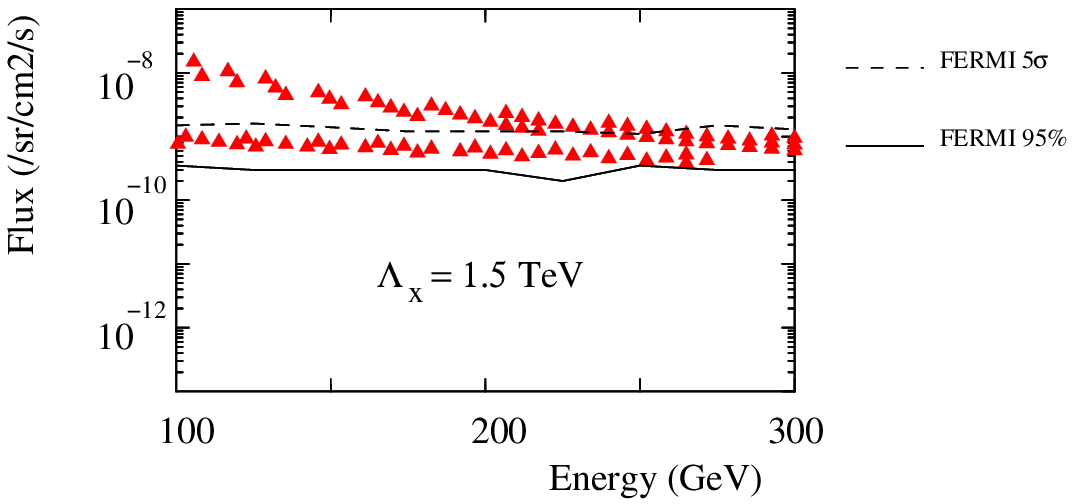}
          \caption{{\footnotesize
Left: example of gamma--ray flux respecting WMAP constraint
for a DM mass of 258 GeV. Right: monochromatic $\gamma-$ray fluxes generated by 
anomaly-cancelation mechanism
in comparison with expected $5\sigma$ and $95\%$ CL sensitivity contours
(5 years of FERMI operation) for the conventional background and unknown WIMP energy,
for an effective scale $\Lambda_X=1.5$ TeV
}}
\label{fig:spectrum}
\end{center}
\end{figure}

As an illustrative point, we show in the left panel of Fig.\ref{fig:spectrum}
an example of  spectrum from the centre annulus that could be observable by the
FERMI telescope, generated by DM annihilation within the pole region
respecting WMAP constraint(
 $m_{\psi_0} = 258$ GeV and $M_{Z_D}=591$ GeV).
We can clearly distinguish a $\gamma-$ray line centered
around $E_\gamma= m_{\psi_0} \left( 1- \frac{M_Z^2}{4 m_{\psi_0}^2} \right)$
above the continuous flux produced by
the annihilation process $\psi_{0} \psi_{0} \rightarrow Z Z / Z \gamma$.
The expected sensitivity of FERMI telescope after 5 years of data taking is
presented in  
the right panel of Fig.\ref{fig:spectrum}.

We clearly see in the right panel of Fig.\ref{fig:spectrum}
that for an effective scale $\Lambda_X = 1.5$ TeV (scale of the "new physics" corresponding
to the fermions generating the anomalies), all the parameter space would be observable
by FERMI at 95\% CL. Indeed, the points that respect the WMAP
constraints lie around the p\^ole $M_{Z_D} \sim 2 m_{\psi_0}$ where
$\sim 60$\% of the annihilation rate is dominated by the $Z\gamma$ final state.
This proportion still holds for annihilating DM in the Galactic
halo and gives a monochromatic line observable by FERMI.

\section{Conclusion}
We showed that the existence of a $dark$
$U(1)_D$ gauge sector which interacts with the Standard Model only through 
its kinetic mixing or anomaly-generated couplings possesses a valid dark matter
 candidate respecting
accelerator, cosmological and the more recent direct detection constraints.
Moreover, considering the latest results of DAMA/LIBRA, CoGENT and 
CRESST, we demonstrated that a specific range of the kinetic mixing 
($\delta \sim  10^{-4}-10^{-3}$) 
can explain all these excesses for a dark boson mass
$M_{Z_D} \sim 10-20 $ GeV, whereas anomaly cancelationconditions 
generate a monochromatic
 $\gamma-$ray line from DM annihilation into $Z \gamma$.
  Such a signature would be a smoking gun signal for
 these types of constructions
It is interesting to notice that other constraints, coming from synchrotron radiation
\cite{Synchrotron}  or difuse gamma-ray emission \cite{Arina:2010rb}
can give more restriction to the analysis.

\section*{Acknowledgements}
Y.M. wants to thank particularly E. Dudas, T. Schwetz, G. Belanger, N.
Fornengo and A. Romagnoni for useful discussions. The work was
supported by the french ANR TAPDMS {\bf ANR-09-JCJC-0146} 
and the spanish MICINNÕs Consolider-Ingenio 2010 Programme 
under grant  Multi- Dark {\bf CSD2009-00064} and
 the E.C. Research Training Networks under contract {\bf MRTN-CT-2006-035505}.
 The author also wants to send all his regards to the "magic" organizing comitee.

\end{document}